\documentclass[12pt]{article}
\begin{document}

\begin{center}
{ \Large Transmutation of Scale Dependence into Truncation Uncertainty
via RG-Improvement of the $R(s)$ Series}
\end{center}

\begin{center}
V. {Elias $^{a,b,}$}\footnote{Electronic address: {\tt velias@uwo.ca}},
D.G.C. {McKeon $^{b}$}\footnote{Electronic address: {\tt dgmckeo2@uwo.ca}}
and T.G. {Steele $^{c}$}\footnote{Electronic address: {\tt Tom.Steele@usask.ca}}
\end{center}

\baselineskip=12pt

\vskip .5cm

\begin{center}
{\small \it
$^{a}$
Perimeter Institute for Theoretical Physics, 35 King Street North,\\
Waterloo, Ontario  N2J 2W9 CANADA}\\
{\small \it
$^{b}$
Department of Applied Mathematics, The University of Western Ontario,\\
London, Ontario  N6A 5B7 CANADA}\\
{\small \it
$^{c}$
Department of Physics and Engineering Physics, University of  Saskatchewan,\\
Saskatoon, Saskatchewan  S7N 5E2  CANADA}
\end{center}

\begin{abstract}
The arbitrariness in how the logarithm is defined within the QCD series for the inclusive
electroproduction cross-section is shown to affect the summation to all
orders in $\alpha_s$ of leading and successively-subleading logarithms
within that perturbative series, even though such summations largely
eliminate the residual dependence of the original series on the
arbitrary renormalization scale $\mu$. However, given that the original
(unimproved) series is known to third-order in $\alpha_s (\mu)$, this 
logarithm ambiguity is
shown not to enter the optimally improved summation-of-logarithms series
until the term {\it fourth-order} in $\alpha_s (s)$, where $s$ is the
physical center-of-mass energy squared.  Consequently, the ambiguity in
how the logarithm is defined is absorbable in the uncertainty associated
with truncating the original perturbative series after its
calculationally known terms.
\end{abstract}

\newpage

\setcounter{footnote}{0}

\baselineskip=24pt

The renormalization-scale ($\mu$) dependence of a perturbative series for a 
physical process necessarily arises as part of the process by which order-by-order
infinities are parametrised and excised.  However, such scale dependence within the 
known terms of a series is widely regarded to be a reflection of the
next-order uncertainty of that series.  This folkloric assertion appears
reasonable insofar as the series taken to {\it all} orders of
perturbation theory must be independent of the unphysical scale $\mu$.
Indeed, the invariance of physical processes under changes in $\mu$ is
the underlying justification for the renormalization-group equation.

However, it is one thing to realize that $\mu$-dependence of a physical
series disappears as the number of series terms increases, but quite
another to say that the $\mu$-dependence exhibited at a given order of
perturbation theory is indicative of truncation uncertainty associated
with our ignorance of the next-order term.  In the present note, we
demonstrate that this stronger assertion is indeed correct within the
context of $\overline {MS}$ QCD corrections to the inclusive
electroproduction cross-section series $R(s)$.  Specifically, we
demonstrate how renormalization-group (RG) improvement of the terms
within the known perturbative series for this process replaces 
$\mu$-dependence with a new ambiguity contingent upon how the 
perturbation-theory logarithm is defined, and that this ambiguity can 
be back-converted into arbitrariness in any RG-optimal choice for the scale
$\mu$.  We are able to show, however, that this new ambiguity does not
enter the unimproved $\mu^2 = s$ perturbative series until the 
(RG-inaccessible) unknown next-order term, reflective of the truncation
uncertainty of the original series.

Radiative corrections to the inclusive electroproduction cross-section
are scaled by a perturbative QCD series ($S$):
\begin{equation}
R(s) \equiv \frac{\sigma(e^+ e^- \rightarrow {\rm hadrons})}{\sigma(e^+ e^-
\rightarrow \mu^+ \mu^-)} = 3 \sum_f Q_f^2 S.
\label{r_def}
\end{equation}
Neglecting quark masses, the perturbative series within Eq.\ (1) is of the form
\begin{equation}
S = 1 + \sum_{n=1}^\infty x^n (\mu) \sum_{m=0}^{n-1} T_{n,m} L^m,
\end{equation}
where the n$^{th}$ power of the QCD couplant $x(\mu) \equiv \alpha_s
(\mu) / \pi$ is multiplied by a degree $n-1$ polynomial in the logarithm
$L \equiv \log (\mu^2 / s)$.  The parameter $\mu$ is the arbitrary
renormalization scale that enters as a consequence of the perturbative
renormalization procedure.  The coefficients $T_{n,0}$ have been
calculated explicitly in the chiral limit from $\overline{MS}$
perturbation theory out to $n = 3$ \cite{1,2},
\footnote{Although the form of (\protect\ref{r_def}) suggests that the singlet contributions proportional to
$x^3\left(\sum_f Q_f\right)^2$  \protect\cite{chetyrkin} have been omitted, such singlet contributions have been absorbed into
the values for $T_{3,0}$.
}
\begin{eqnarray}
n_f=3: && \; \; \; T_{1,0} = 1, \; \; \; T_{2,0} = 1.63982, \; \; \;
T_{3,0} = -10.2839,\nonumber\\
n_f=4: && \; \; \; T_{1,0} = 1, \; \; \; T_{2,0} = 1.52453, \; \; \;
T_{3,0} = -11.6856,\nonumber\\
n_f=5: && \; \; \; T_{1,0} = 1, \; \; \; T_{2,0} = 1.40924, \; \; \;
T_{3,0} = -12.8046,
\end{eqnarray}
and for $n \leq 3$, the coefficients $T_{n,m}$ ($1 \leq m \leq n-1$) can
easily be obtained from the process-appropriate RG equation \cite{1,3}:
\begin{equation}
T_{2,1} = \beta_0, \; \; \; T_{3,1} = 2\beta_0 T_{2,0} + \beta_1, \; \;
\; T_{3,2} = \beta_0^2,
\end{equation}
where \cite{4}
\begin{eqnarray}
\beta_0 = (11 - 2n_f / 3)/4, \; \; \beta_1 = (102 - 38 n_f / 3)/16,
\nonumber\\ \beta_2 = (2857/2 - 5033n_f / 18 + 325 n_f^2 / 54)/64.
\end{eqnarray}
The form of the series (2) is unaffected by a redefinition of the
logarithm.  If we redefine the logarithm to be
\begin{equation}
L_k \equiv \log (k\mu^2/s) = L - \log(k),
\end{equation}
the series (2) becomes
\begin{equation}
S = 1 + \sum_{n=1}^\infty x^n (\mu) \sum_{m=0}^{n-1}\tilde{T}_{n,m}
(L_k)^m,
\end{equation}
where
\begin{eqnarray}
&& \tilde{T}_{1,0} = T_{1,0} \; (=1), \; \; \tilde{T}_{2,0} = T_{2,0} - T_{2,1} \log
(k), \; \; \tilde{T}_{2,1} = T_{2,1},\nonumber\\
&& \tilde{T}_{3,0} = T_{3,0} - T_{3,1} \log (k) + T_{3,2} \log^2
(k),\nonumber\\
&& \tilde{T}_{3,1} = T_{3,1} - 2 T_{3,2} \log (k), \; \; \tilde{T}_{3,2}
= T_{3,2}.
\end{eqnarray}
Thus the redefinition (6) of the logarithm is compensated trivially by
appropriate shifts in the coefficients $T_{m,n}$, such that the series
(2) and (7) agree order-by-order in the expansion parameter $x(\mu)$.
Note that incorporation of the redefined logarithm $L_k$ (6) does not
entail any compensating change in the renormalization scale $\mu$;  the
argument of $x(\mu)$ has not been altered.

In refs.\ \cite{5,6}, it is shown that truncations of the series (2) can
be optimally RG-improved through inclusion of all terms in that series
accessible via the RG-equation, an approach suggested first (to our
knowledge) by Maxwell \cite{7}.  Given the current determination of the
$\overline{MS}$ QCD $\beta$-function to four orders in $\alpha_s$
\cite{8}, one can show that knowledge of $T_{1,0}$ is sufficient to
determine all leading-logarithm coefficients $T_{n, n-1}$;  knowledge of
$T_{2,0}$ is sufficient to determine all next-to-leading logarithm
coefficients $T_{n, n-2}$;  and knowledge of $T_{3,0}$ is sufficient to
determine all two-from-leading logarithm coefficients $T_{n, n-3}$
\cite{5}.  Thus one can obtain {\it all-orders} summations of
sequentially-subleading logarithms within the series (2):
\begin{eqnarray}
S \rightarrow S_\Sigma & = & 1 + x(\mu) \sum_{n=1}^\infty T_{n, n-1}
\left( x(\mu) L\right)^{n-1}\nonumber\\
& + & x^2 (\mu) \sum_{n=2}^\infty T_{n, n-2} \left( x(\mu) L\right)^{n-
2}\nonumber\\
& + & x^3 (\mu) \sum_{n=3}^\infty T_{n, n-3} \left( x(\mu) L \right)^{n-
3} + ...\nonumber\\
& = & 1 + x(\mu) S_1 \left( x(\mu) L \right) + x^2 (\mu) S_2 \left(
x(\mu) L \right) + x^3 (\mu) S_3 \left( x(\mu) L \right) +
...\nonumber\\
\end{eqnarray}
In ref.\ \cite{6}, $S_1$, $S_2$ and $S_3$ are calculated explicitly from
their generating coefficients $T_{1,0}$, $T_{2,0}$, $T_{3,0}$ by solving
successive differential equations obtained by inserting Eq.\ (9) into the
RG equation.  The solutions obtained in ref.\ \cite{6} are
\begin{equation}
S_1 (xL) = \frac{1}{(1 - \beta_0 xL)}
\end{equation}
\begin{equation}
S_2 (xL) = \frac{T_{2,0} - \frac{\beta_1}{\beta_0} \log \left( 1 - \beta_0
xL \right)}{\left( 1-\beta_0 xL \right)^2},
\end{equation}
\begin{eqnarray}
S_3 (xL) & = & \left( \frac{\beta_1^2}{\beta_0^2} - \frac{\beta_2}{\beta_0}
\right) / \left( 1 - \beta_0 xL \right)^2 \nonumber\\
& + & \frac{\left\{T_{3,0} - \left( \frac{\beta_1^2}{\beta_0^2} - \frac{\beta_2%
}{\beta_0} \right) - \frac{\beta_1}{\beta_0} \left( 2 T_{2,0} + \frac{\beta_1%
}{\beta_0} \right) \log (1 - \beta_0 xL) + \frac{\beta_1^2}{\beta_0^2}
\log^2 (1 - \beta_0 xL) \right\} } {\left( 1 - \beta_0 xL
\right)^3}.\nonumber\\
\end{eqnarray}
Since the infinite series (9) is by construction independent of the
renormalization scale $\mu$, it is not surprising that truncations
of the RG-improved series (9) exhibit far less dependence on the 
renormalization scale $\mu$ than corresponding truncations of the 
original series (2), which necessarily reflect residual 
renormalization-scale dependence \cite{5, 6}.  For example, if the
series (2) is truncated after its known ${\cal{O}}(x^3)$ contributions,
one finds for $n_f = 5$ and $\sqrt{s} = 15$ GeV that this truncated
unimproved series ($S^{(3)}$) increases from $1.0525$ to $1.0540$ as
$\mu$ varies from $\sqrt{s}/2$ to $2\sqrt{s}$, given 4-loop evolution of
$\alpha_s (\mu)$ from $\alpha_s(M_z) = 0.11800$ \cite{6}.  By contrast,
a corresponding truncation $S_{\Sigma}^{(3)}$ of the optimally RG-improved
series (9) after its $x^3(\mu) S_3 (x(\mu) L)$ term eliminates almost
all such residual scale dependence;  {\it i.e.}, $S_\Sigma^{(3)}$ stays
between $1.0537$ and $1.0538$ over the same range of $\mu$ \cite{6}.
Since Eqs.\ (2) and (9) are identical when $L = 0$, one can argue that
such optimal RG-improvement favours the physical choice $\mu = \sqrt{s}$
for the renormalization scale occurring in the original series (2).

However, this elimination of scale dependence appears to
come at a price.  Suppose one redefines the logarithm via Eq.\ (6) and
then sums successively-subleading logarithms in the series (7).  One
finds that this new summation-of-logarithms series {\it differs} from the
series (9), even though the unimproved series (2) and (7) are 
order-by-order equivalent.  The new summation-of-logarithms series is 
\begin{eqnarray}
S \rightarrow S_{\Sigma^\prime} & = & 1 + x(\mu) \tilde{S}_1 \left(
x(\mu) L_k \right) + x^2 (\mu) \tilde{S}_2 \left( x (\mu) L_k
\right)\nonumber\\
& + & x^3 (\mu) \tilde{S}_3 \left( x(\mu) L_k \right) + ...
\end{eqnarray}
where
\begin{equation}
\tilde{S}_1 (x L_k) = \frac{1}{(1 - \beta_0 x L_k)}
\end{equation}
\begin{equation}
\tilde{S}_2 (x L_k) = \frac{\tilde{T}_{2,0} - \frac{\beta_1}{\beta_0} \log \left( 1 - \beta_0
x L_k \right)}{\left( 1-\beta_0 x L_k \right)^2},
\end{equation}
\begin{eqnarray}
\tilde{S}_3 (x L_k) & = & \left( \frac{\beta_1^2}{\beta_0^2} - \frac{\beta_2}{\beta_0}
\right) / \left( 1 - \beta_0 x L_k \right)^2 \nonumber\\
& + & \frac{\left\{\tilde{T}_{3,0} - \left( \frac{\beta_1^2}{\beta_0^2} - \frac{\beta_2%
}{\beta_0} \right) - \frac{\beta_1}{\beta_0} \left( 2 \tilde{T}_{2,0} + \frac{\beta_1%
}{\beta_0} \right) \log (1 - \beta_0 x L_k) + \frac{\beta_1^2}{\beta_0^2}
\log^2 (1 - \beta_0 x L_k) \right\} } {\left( 1 - \beta_0 x L_k
\right)^3}.\nonumber\\
\end{eqnarray}
For example, if one truncates both unimproved series (2) and (7) after only two terms 
and then notes [Eq.\ (8)] that $T_{1,0} = \tilde{T}_{1,0} = 1$, one finds that
corresponding truncations of the summation-of-leading logarithms series
are inequivalent because of their differing logarithms:
\begin{equation}
S_\Sigma^{(1)} = 1 + x (\mu) / \left( 1 - \beta_0 x (\mu) L \right),
\end{equation}
\begin{equation}
S_{\Sigma^\prime}^{(1)} = 1 + x (\mu) / \left( 1 - \beta_0 x (\mu) L_k
\right).
\end{equation}
If both expressions are comparably free of residual $\mu$-dependence, we
see that the $\mu$-dependence from the corresponding truncation of the
unimproved series,
\begin{equation}
S^{(1)} = 1 + x(\mu),
\end{equation}
appears to be replaced by the $k$-dependence in Eq.\ (18), a reflection
of the arbitrariness in how the logarithm is defined.  Moreover, the
(essentially $\mu$-independent) RG-improved series (13) coincides with
the unimproved series (7) at $L_k = 0$, corresponding in this latter
series to a preferred value for $\mu$ of $\mu = \sqrt{s/k}$, {\it where
$k$ is arbitrary}. \footnote{VE is grateful to P. Lepage for pointing 
out this ambiguity, which is also discussed in ref.\ \cite{9}.}  Thus
the ambiguity in the renormalization-scale $\mu$ chosen for the
unimproved series is replaced by a corresponding ambiguity in how the
logarithms are to be {\it defined} in the improved series prior to their
summation.

In actual fact, such $k$-dependence can be subsumed entirely into the
truncation uncertainty of the Eq.\ (19) when $\mu$ is chosen to be the
external physical scale $\sqrt{s}$.  Thus, we argue that the 
$k$-dependence of Eq.\ (18) is really a reflection of the error implicit 
in truncating the series (2) after its first two terms (19).  To see 
this, consider first the explicit $\mu$-dependence of the one loop 
running couplant ($\mu^2 dx/d\mu^2 = -\beta_0 x^2$) about its 
RG-invariant reference value $x(\sqrt{s})$:
\begin{equation}
x(\mu) = x (\sqrt{s}) / \left( 1 + \beta_0 x (\sqrt{s}) \log (\mu^2 / s)
\right).
\end{equation}
If we substitute Eq.\ (20) into Eq.\ (18), we find that
\begin{eqnarray}
S_{\Sigma^\prime}^{(1)} & = & 1 + x (\sqrt{s}) / \left( 1 - \beta_0 x
(\sqrt{s}) \log (k) \right)\nonumber\\
& = & 1 + x (\sqrt{s}) + x^2 (\sqrt{s}) \beta_0 \log (k) + ...
\end{eqnarray}
Note that this expression is totally independent of the renormalization
scale $\mu$.  Moreover, its dependence upon $\log (k)$ [the logarithm
ambiguity] occurs only in the next order of $x (\sqrt{s})$,
{\it i.e.}, within a term not determined by the RG equation.
Consequently, Eq.\ (21) is fully consistent with the unimproved
expression (19) when $\mu = \sqrt{s}$, {\it regardless of $k$}.

Surprisingly, we find that the transfer all $k$-dependence to the
first post-truncation order (as well as the absence of any 
$\mu$-dependence in the first post-truncation order as well 
as in previous orders) is upheld when the next
two subsequent orders of perturbation theory are taken into account.
Suppose one wishes to make an optimal RG-improvement of the series $S$
[Eq.\ (7)] truncated after all of its known terms:
\begin{eqnarray}
S^{(3)} & = & 1 + x (\mu) + \left( \tilde{T}_{2,0} + \tilde{T}_{2,1} L_k
\right) x^2 (\mu)\nonumber\\
& + & \left( \tilde{T}_{3,0} + \tilde{T}_{3,1} L_k + \tilde{T}_{3,2} L_k^2
\right) x^3 (\mu).
\end{eqnarray}
As before, one can express the running couplant $x(\mu)$ in Eq.\ (22) in
terms of the physical reference value $x(\sqrt{s})$.  In Eq.\ (5.14) of
ref.\ \cite{10}, summation of logarithm techniques are employed to expand
any RG-invariant effective couplant $x(p)$ in powers of the running
couplant $x(\mu)$.  A straightforward inversion of this expression with
$p = \sqrt{s}$ yields the following three leading terms:
\begin{eqnarray}
&& x^{(3)} (\mu) = x(\sqrt{s})\left[ \frac{1}{1 + \beta_0 \; x(\sqrt{s}) L}\right]\nonumber\\
& - & x^2 (\sqrt{s}) \left[ \frac{\beta_1 \log [1 + \beta_0 \; x(\sqrt{s}) L]}{\beta_0 [1+\beta_0 \; x (\sqrt{s}) L]^2} \right]\nonumber\\
& + & x^3 (\sqrt{s}) \left[ \frac{(\beta_1^2 - \beta_2 \beta_0) \beta_0 x (\sqrt{s}) L - 
\beta_1^2 \left\{ \log \left[ 1 + \beta_0 x (\sqrt{s}) L \right] 
- \log^2 \left[ 1 + \beta_0 x(\sqrt{s}) L \right] \right\}}{\beta_0^2 \left[ 1 + \beta_0 x (\sqrt{s}) L \right]^3} \right],\nonumber\\
\end{eqnarray}
where $L \equiv log (\mu^2 / s)$, as before. If one substitutes Eq.\ (23) for 
$x(\mu)$ everywhere it appears within the RG-improvement of Eq.\ (22),
\begin{eqnarray}
S_{\Sigma^\prime}^{(3)} & = & 1 + x(\mu) \tilde{S}_1 \left( x (\mu) L_k
\right) + x^2 (\mu) \tilde{S}_2 \left( x(\mu) L_k \right)\nonumber\\
& + & x^3 (\mu) \tilde{S}_3 \left( x(\mu) L_k \right),
\end{eqnarray}
one obtains the following power series expansion in $x(\sqrt{s})$ via Eqs.
(8) and (14-16):
\begin{eqnarray}
S_{\Sigma^\prime}^{(3)} & = & 1 + x(\sqrt{s}) + T_{2,0} x^2 (\sqrt{s}) +
T_{3,0} x^3 (\sqrt{s})\nonumber\\
& + & x^4 (\sqrt{s}) \biggl[ 2\beta_0^3 \log^2 (k) - \left( 6\beta_0^2
T_{2,0} + 5 \beta_1 \beta_0 \right) \log (k) \biggr.\nonumber\\
& + & \biggl. 6\beta_0 T_{3,0} + 4\beta_1 T_{2,0} + 2\beta_2 \biggr] \log
(k)/2\nonumber\\
& + & x^5 (\sqrt{s}) \left[ A(k) + B(k) \log (\mu^2 / s) \right]
+{\cal O}\left(x^6\right).
\end{eqnarray}
where
\begin{eqnarray}
A(k) & = &  \biggl[ 3\beta_0^4 \log^2 k - \left( 8 T_{2,0} \beta_0^3 +
\frac{26}{3} \beta_0^2 \beta_1 \right) \log k \biggr.\nonumber\\
& + & \biggl. 7 \beta_1 \beta_0 T_{2,0} + 3 \beta_2 \beta_0 + 6 T_{3,0}
\beta_0^2 \biggr] \log^2 k,\nonumber\\
B(k) & = & - 3\beta_0^2 \log^2 k + \left( 3\beta_1^2 + 6 \beta_0 \beta_1
T_{2,0} + 2 \beta_2 \beta_0 \right) \log k\nonumber\\
& - & 3 \beta_1 T_{3,0} - 2 \beta_2 T_{2,0}.
\end{eqnarray}
The power series (25) is consistent with Eq.\ (22), the ${\cal{O}}(x^3)$
truncation of the unimproved series (7) evaluated at $\mu = \sqrt{s}$.
The $k$-ambiguity ( {\it i.e.} the arbitrariness in how the logarithm is
defined) is entirely absorbed in the not-yet-known next order of
perturbation theory;  indeed this ${\cal{O}}\left( x^4
(\sqrt{s})\right)$ contribution is zero if $k$ is chosen equal to one.
Moreover, this contribution does not exhibit any renormalization scale
dependence, which is not seen to arise until the ${\cal{O}}\left( x^5
(\sqrt{s})\right)$ contribution to Eq.\ (25). In other words, neither the
$k$-ambiguity nor any $\mu$-dependence occurs in terms that should be
determined by the RG equation.  The parameter $k$ does not occur until the
first RG-inaccessible order, and the parameter $\mu$ does not occur until the
second RG-inaccessible order.

Thus, if one implements RG-improvement on Eq.\ (22), which is
just the series (7) truncated after calculationally known terms, by
summing leading and two subsequently subleading sets of logarithms to
all orders of perturbation theory, one finds that the arbitrariness in
how the logarithm in the series (7) is defined does not enter the
RG-improved expression (25) until the first post-truncation order of
$x(\sqrt{s})$, a term sensitive to the not-yet-known coefficient
$T_{4,0}$.  Consequently, the ambiguity in how the logarithm is defined can be absorbed
in the truncation-uncertainty of the series.  This uncertainty is
decoupled from any residual dependence on the renormalization scale parameter $\mu$.
The RG-improved expression (25) is seen to retain renormalization-scale independence
even to this first post-truncation order.

In the truncation of a conventional perturbative series such as Eq.\ (2), minimization of the residual scale dependence is known to be
of value in extracting information about higher order terms.  Such an approach has, for example, been employed in the extraction of
$\alpha_s$ from deep inelastic scattering structure functions \cite{kataev1}.  Pertinent to our present analysis, the minimization of
residual scale dependence has also proved useful in obtaining estimates of the (as-yet-uncalculated) series (2) coefficient $T_{4,0}$
\cite{kataev2}, estimates which appear to be corroborated by the ${\cal O}(n_f^2)$ contributions to the absorptive parts of the
five-loop vector-current vacuum polarization function  \cite{baikov} used to construct $R(s)$.
The values of $\log(k)$ that respectively correspond
to ref.\ \cite{kataev2}'s
predicted values for $T_{4,0}$ [-128 ($n_f=3$), -112 ($n_f=4$), -97 ($n_f=5$)] are $\log k=\{-1.76, 1.41, 4.51\}$
for $n_f=3$, $\log k=\{-2.03, 1.26, 4.82\}$ for $n_f=4$, and $\log k=\{-2.38,1.13 , 5.10\}$ for $n_f=5$.

Within Eq.\ (25), one might similarly speculate that minimization of the sensitivity of the $x^4(\sqrt{s})$ coefficient to changes in
$k$ might also serve to predict $T_{4,0}$, or at least its approximate magnitude, consistent
with Stevenson's Principle of Minimal Sensitivity \cite{stevenson}.
If one optimizes Eq.\ (25)'s $x^4(\sqrt{s})$ coefficient with respect to $\log(k)$, one obtains a
quadratic equation whose solution yields two real optimization points.  The larger of these points of minimal sensitivity to $k$ yields
values with the same sign and approximately double the magnitude of ref.\ \cite{kataev2} predictions for $T_{4,0}$.
However, values of $\log(k)$ for which $A(k)=0$, eliminating truncation uncertainty at ${\cal O}(x^5)$ for $\mu=\sqrt{s}$,
respectively result in the predictions $T_{4,0}=\{-117,-126,-133\}$ for $n_f=\{3,4,5\}$--- values  in reasonable
agreement with the ref.\ \cite{kataev2} estimates.
It is particularly of interest that such $T_{4,0}$ estimates were obtained via the Adler function in the Euclidean
momentum region \cite{kataev2}, and that they nevertheless appear to be corroborated by the Minkowski region approach
delineated above.  Moreover,
different choices of renormalization scale do not have a significant effect on the
values of $\log(k)$ that eliminate the full ${\cal O}(x^5)$ truncation uncertainty.

To summarize, residual renormalization-scale dependence of truncations
of the perturbative series (7) [which is term-by-term equivalent to the
original series (2)] no longer occurs within the optimally RG-improved
series obtained through all-orders summation of leading and
successively-subleading logarithms \cite{6}.  However, such RG-improved
results now depend on how the logarithm is defined, {\it i.e.}, the parameter
$k$ in Eq.\ (6), even though truncations of the unimproved series (7) are
independent of this choice.  Moreover, a clear correspondence ($\mu =
\sqrt{s/k}$) exists between the ``k-ambiguity'' of the improved series
and the scale $\mu$-dependence of the unimproved series.  We have shown here
that the ambiguity in how the logarithm is defined is ultimately a
reflection of the uncertainty deriving from truncation of the series
itself.  Thus, residual renormalization-scale dependence is {\it
demonstrably} a reflection of the uncertainty associated with the
unknown next-order term.

We are grateful for research support from the Natural Sciences and Engineering
Research Council of Canada (NSERC), and for discussions with  A.\ Rebhan and F.A.\ Chishtie.

\end{document}